\documentclass[%
 aip,
 amsmath,amssymb,
 reprint,%
]{revtex4-1}

\usepackage{graphicx}
\usepackage{dcolumn}
\usepackage{bm}
\usepackage{makecell} 
\usepackage{booktabs} 

\usepackage[utf8]{inputenc}
\usepackage[T1]{fontenc}
\usepackage{mathptmx}
\usepackage{siunitx}
\usepackage[version=4]{mhchem}

\usepackage{hyperref}
\hypersetup{
    colorlinks=true,
    linkcolor=blue,
    filecolor=blue,      
    urlcolor=blue,
    citecolor=blue,
}

\usepackage{color,soul}
\usepackage{xspace}
\usepackage{xcolor}
\usepackage{tablefootnote}

\makeatletter
\def\@email#1#2{%
 \endgroup
 \patchcmd{\titleblock@produce}
  {\frontmatter@RRAPformat}
  {\frontmatter@RRAPformat{\produce@RRAP{*#1\href{mailto:#2}{#2}}}\frontmatter@RRAPformat}
  {}{}
}%
\makeatother
\begin{document}

\preprint{AIP/123-QED}

\title{Thermodynamic Modeling of Fluid Polyamorphism in Hydrogen at Extreme Conditions}

\author{Nathaniel R. Fried}
\affiliation{Institute for Physical Science and Technology, University of Maryland, College Park, MD 20742, USA}

\author{Thomas J. Longo}
\affiliation{Institute for Physical Science and Technology, University of Maryland, College Park, MD 20742, USA}


\author{Mikhail A. Anisimov}
\affiliation{Institute for Physical Science and Technology, University of Maryland, College Park, MD 20742, USA}
\affiliation{Department of Chemical and Biomolecular Engineering, University of Maryland, College Park, MD 20742, USA}
\email{tlongo1@umd.edu, anisimov@umd.edu}


\date{\today}
\begin{abstract}
Fluid polyamorphism, the existence of multiple amorphous fluid states in a single-component system, has been observed or predicted in a variety of substances. A remarkable example of this phenomenon is the fluid-fluid phase transition in high-pressure hydrogen between insulating and conducting high-density fluids. This transition is induced by the reversible dimerization/dissociation of the molecular and atomistic states of hydrogen. In this work, we present the first attempt to thermodynamically model the fluid-fluid phase transition in hydrogen at extreme conditions.  Our predictions for the phase coexistence and the reaction equilibrium of the two alternative forms of fluid hydrogen are based on experimental data and supported by the results of simulations. Remarkably, we find that the law of corresponding states can be utilized to construct a unified equation of state combining the available computational results for different models of hydrogen and the experimental data.
\end{abstract}

\maketitle

In addition to being a liquid or a gas, single-component substances can exist in other amorphous fluid states. This phenomenon is known as liquid or, more generally, fluid polyamorphism \cite{Morales_H_2010,Sciorino_Silicon_2011,Stanley_Liquid_2013,Anisimov_Polyamorphism_2018,Tanaka_Liquid_2020}. Fluid polyamorphism has been observed or predicted in a variety of substances, such as superfluid helium \cite{Vollhardt_He_1990,Schmitt_He_2015}, high-pressure-fluid hydrogen \cite{Ohta_H_2015,Zaghoo_H_2016,McWilliams_H_2016,Norman_Review_2021}, sulfur \cite{Henry_Sulfur_2020}, phosphorous \cite{Katayama_Phos_2000,Katayama_Phos_2004}, liquid carbon \cite{Glosli_Liquid_1999}, silicon \cite{Sastry_Silicon_2003,Beye_Silicon_2010,Vasisht_Solicon_2011}, silica \cite{Saika_Silica_2000,Lascaris_Silica_2014}, selenium and tellurium \cite{Tsuchiya_SeTe_1982,Brazhkin_SeTe_1999}, and cerium \cite{Cadien_Ce_2013}. It is also highly plausible to exist in metastable deeply supercooled liquid water below the temperature of spontaneous ice nucleation \cite{Angell_TwoState_1971,Angell_Amorphous_2004,Stanley_Liquid_2013,Anisimov_Polyamorphism_2018,Tanaka_Liquid_2020,Poole_Water_1992,Debenedetti_Water_1998,Holten_Water_2012,Holten_Water_2014,Gallo_Water_2016,Biddle_Water_2017,Caupin_Thermodynamics_2019,Duska_Water_2020}. 

Fluid polyamorphism can be modeled thermodynamically through the reversible interconversion of two alternative molecular or supramolecular states \cite{Anisimov_Polyamorphism_2018,Caupin_Polyamorphism_2021,Longo_PT_2022}. The application of this ``two-state'' thermodynamics to the variety of polyamorphic substances could be just as useful a phenomenology that may or may not necessarily reflect the microscopic origin of polyamorphism. However, there are a few substances, such as hydrogen, sulfur, phosphorous, and carbon, where the existence of alternative liquid or dense-fluid states can be explicitly induced by a reversible chemical reaction: polymerization in sulfur, phosphorus, and carbon or dimerization in hydrogen \cite{Shumovskyi_Sulfur_2022}. 

In this work, based on the available experimental and computational information obtained for this phenomenon, we present the first attempt to thermodynamically model the first-order fluid-fluid phase transition between molecular (dielectric) and atomistic (conductive) states of hydrogen. Experiments and simulations have discovered that at extremely high pressures, highly-dense fluid (dimeric) hydrogen dissociates into atomistic fluid hydrogen \cite{Weir_FirstMesur_1996,Tonkov_HighPress_Book,Brazhkin_HighPres_1997,Zaghoo_H_2013,Zaghoo_H_2016,Zaghoo_H_2017,McWilliams_H_2016,Ohta_H_2015,Morales_H_2010,Lorenzen_H_2010,Morales_Review_2012,Scandolo_2013,Morales_LowT_2015,Pierleoni_H_2016,Mazzola_H_2018,Geng_H_2019,Heinz_H_2020,Cheng_H_2020,Cheng_Comment_2021,Norman_Review_2021}. Using the generalized law of corresponding states, by reducing the temperature, pressure, and entropy by their critical values, we combine the available experimental data with the results of computations\cite{Lorenzen_H_2010,Morales_H_2010,Morales_Review_2012,Xu_Supercritical_2015,Pierleoni_H_2016,Mazzola_H_2018,Geng_H_2019,Heinz_H_2020,Cheng_H_2020,Cheng_Comment_2021,Tirelli_MachineLearn_2022} to predict the equation of state of hydrogen near the fluid-fluid phase transition (FFPT). We show predictions for the phase coexistence and the reaction equilibrium of the two alternative states of fluid hydrogen.

There is a remarkable analogy between the challenges in thermodynamic modeling of fluid polyamorphism in hydrogen and that in supercooled water. In both cases, there is a reasonable agreement on the shape and location of the first-order transition line, while the position of the fluid-fluid critical point (FFCP) is highly uncertain and a subject of current debate in the literature \cite{Gallo_Water_2016,Geng_H_2019,Zaghoo_H_2013,Zaghoo_H_2016,Zaghoo_H_2017,Comment_Zaghoo_2016_1,Comment_Zaghoo_2016_2,Reply_Comment_Zaghoo_2016}. This uncertainty, in both hydrogen and water, is  due to the extreme conditions of the phenomena. In supercooled water, the liquid-liquid transition is hidden below the temperature of spontaneous ice formation \cite{Gallo_Water_2016,Debenedetti_Water_1998}, while in hydrogen, the fluid-fluid transition occurs at immensely high pressures (millions of atm) \cite{Geng_H_2019}. Consequently, it is not surprising that the available computational or experimental data are scarce.\cite{Morales_Review_2012,Geng_H_2019,Heinz_H_2020}  We show that despite the uncertainty in determining the location of the FFCP in hydrogen, thermodynamic modeling provides a principle direction to predict the equation of state for the system. Remarkably, we find that the law of corresponding states can be utilized to reconcile the different computational models of hydrogen and experiment \cite{Zaghoo_H_2013,Zaghoo_H_2016,Zaghoo_H_2017,McWilliams_H_2016,Ohta_H_2015,Morales_H_2010,Lorenzen_H_2010,Morales_Review_2012,Xu_Supercritical_2015,Pierleoni_H_2016,Mazzola_H_2018,Geng_H_2019,Heinz_H_2020,Cheng_H_2020,Cheng_Comment_2021} into a unified equation of state. We introduce an additional parameter to generalize the law of corresponding states, the entropy at the critical point ($S_\text{c}$), which provides the opportunity for further studies of hydrogen, both experimental and computational, to be unified under the general approach presented in this work. 

\begin{figure*}[t]
    \centering
    \includegraphics[width=0.49\linewidth]{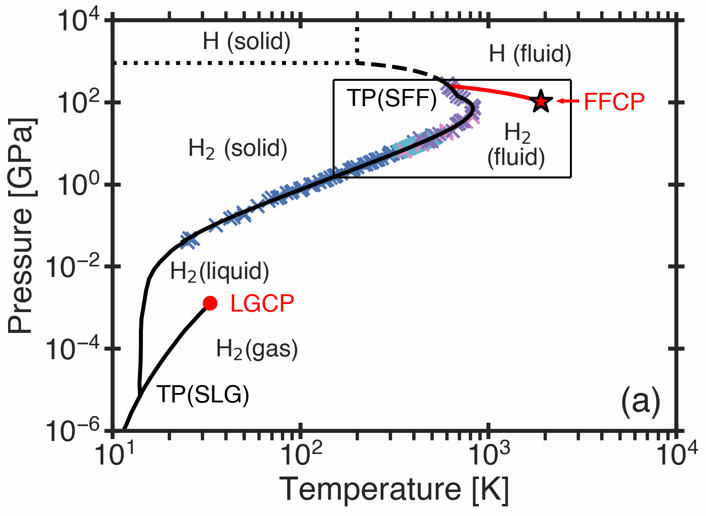}
    \includegraphics[width=0.49\linewidth]{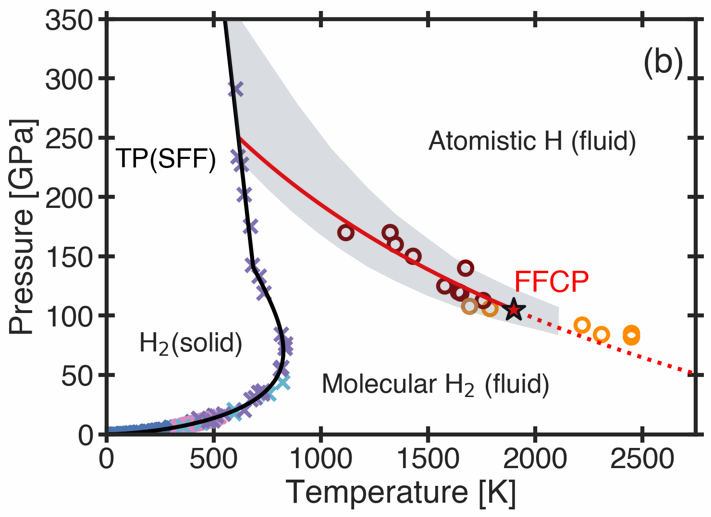}
    \caption{The global pressure-temperature phase diagram for hydrogen. (a) The full range from low to extreme pressures in logarithmic scale. The crosses indicate the experimental data for the solid-liquid melting transition presented in Diatschenko \textit{et al.} \cite{Diatschenko_MeltExp_1985} (blue), Datchi \textit{et al.} \cite{Datchi_MeltExp_2000} (cyan), Gregoryanz \textit{et al.} \cite{Gregoryanz_MeltExp_2003} (pink), and Zha \textit{et al.} \cite{Zha_MeltExp_2017} (purple). The solid black curves at low pressure ($P\le \SI{0.1}{\giga\pascal}$) are the liquid-gas-solid phase transitions \cite{Fukai_H_Properties}, while the solid black curve at high pressure  ($P > \SI{0.1}{\giga\pascal}$) is the Kechin equation \cite{Kechin_Equation_2001} as reported in ref. \cite{Zha_MeltExp_2017}. The black dashed curve is the predicted continuation of the melting line based on experimental and computational evidence \cite{Simpson_H_2016,Zha_MeltExp_2017,Cheng_H_2020}, while the dotted lines represent the highly-debated prediction \cite{Gregoryanz_SSExp_Review_2020,Geng_Comment_DiasSilvera_2017,Eremets_Comment_DiasSilvera_2017,Loubeyre_Comment_DiasSilvera_2017,Goncharov_Comment_DiasSilvera_2017,Dias_Silvera_Response_2017,Monacelli_QPT_Dia_2022} of the domain of solid metallic hydrogen \cite{Wigner_Huntington_MetalH_1935,Dias_Silvera_SSExp_2017,Dias_Silvera_Review_Hydrogen_2018,Eremets_SSExp_2019,Loubeyre_SSExp_2020,Gregoryanz_SSExp_Review_2020,Geng_Comment_DiasSilvera_2017,Monacelli_QPT_Dia_2022}. The red line is the first-order fluid-fluid phase transition adopted in this work. (b) The phase diagram of hydrogen at extreme conditions, in the area of the box in (a). The open circles are experimental data presented in Zaghoo \textit{et al.} \cite{Zaghoo_H_2013,Zaghoo_H_2016,Zaghoo_H_2017} (dark brown), McWilliams \textit{et al.} \cite{McWilliams_H_2016} (light brown), and Ohta \textit{et al.} \cite{Ohta_H_2015} (orange). Simulation results \cite{Lorenzen_H_2010,Morales_Review_2012,Pierleoni_H_2016,Mazzola_H_2018,Geng_H_2019,Heinz_H_2020,Cheng_H_2020,Cheng_Comment_2021,Tirelli_MachineLearn_2022} are spread within the grey area and shown in detail in Fig.~\ref{Fig_CorrespondStates}. The fluid-fluid phase transition (solid red) and Widom line (dotted red) are represented by Eq.~(\ref{Eq_rxn_cond}). The red star is the location of the fluid-fluid critical point (FFCP) as adopted in this work.}
    \label{Fig_PT}
\end{figure*}

The suggested global phase diagram of hydrogen, based only on the available experimental evidence for the fluid-fluid phase transition \cite{Zaghoo_H_2013,Zaghoo_H_2016,Zaghoo_H_2017,McWilliams_H_2016,Ohta_H_2015}, the solid-liquid melting transition \cite{Diatschenko_MeltExp_1985,Datchi_MeltExp_2000,Gregoryanz_MeltExp_2003,Deemyad_Silvera_MeltExp_2008,Eremets_MeltExp_2009,Subramanian_MeltExp_2011,Zha_MeltExp_2017}, and the location of solid-metallic hydrogen \cite{Wigner_Huntington_MetalH_1935,Dias_Silvera_SSExp_2017,Dias_Silvera_Review_Hydrogen_2018,Eremets_SSExp_2019,Loubeyre_SSExp_2020,Gregoryanz_SSExp_Review_2020}, which is supported by the most recent computational studies\cite{Lorenzen_H_2010,Morales_H_2010,Morales_Review_2012,Xu_Supercritical_2015,Pierleoni_H_2016,Mazzola_H_2018,Geng_H_2019,Heinz_H_2020,Cheng_H_2020,Cheng_Comment_2021,Bonev_MeltSim_2004,Attaccalite_MeltSim_2008,Liu_MeltSim_2012,Belonoshko_MeltSim_2013}, is shown in Fig.~\ref{Fig_PT}a. It illustrates the fact that a huge pressure gap separates the liquid-gas \cite{Fukai_H_Properties} and fluid-fluid phase transitions in hydrogen.

Our adopted locations of the FFCP and the solid-fluid-fluid triple point (SFF-TP) are based on the available experimental data \cite{Zaghoo_H_2013,Zaghoo_H_2016,Zaghoo_H_2017,Ohta_H_2015,McWilliams_H_2016} and on discussions present in the literature \cite{Mazzola_H_2018,Geng_H_2019,Heinz_H_2020} (Table~\ref{Table_Key_Points}). We note that the exact location of the FFCP is uncertain, as the interpretation of both the Zaghoo \textit{et al.} \cite{Zaghoo_H_2016} and Ohta \textit{et al.} \cite{Ohta_H_2015} experimental data have been highly debated \cite{Comment_Zaghoo_2016_1,Comment_Zaghoo_2016_2,Reply_Comment_Zaghoo_2016,Geng_H_2019}. Most authors suggest that the experimental data of Ohta \textit{et al.}\cite{Ohta_H_2015}, on the anomalies of the heating efficiency, are obtained in the supercritical region\cite{Geng_H_2019}. We interpret the results observed by Ohta \textit{et al.}\cite{Ohta_H_2015} as the anomalies of the heating efficiency along the ``Widom line'', the line corresponding to the maximum of the fluctuations of the order parameter, which emanates from the critical point \cite{Gallo_Water_2016,Anisimov_Polyamorphism_2018,Saitov_Metastable_2019}.

\begin{table}[h!]
\caption{The suggested locations of the FFCP and the SFF-TP.}
\begin{tabular}{lccc}
         & $P$ [GPa] & $T$ [K] & $\rho$ [g/cm$^3$] \\ \hline
        FFCP     & 105                & 1900                & 0.8       \\
        SFF-TP & 250                & 600                 & -         
\end{tabular}
\label{Table_Key_Points}
\end{table}

\begin{figure}
    \centering
    \includegraphics[width=\linewidth]{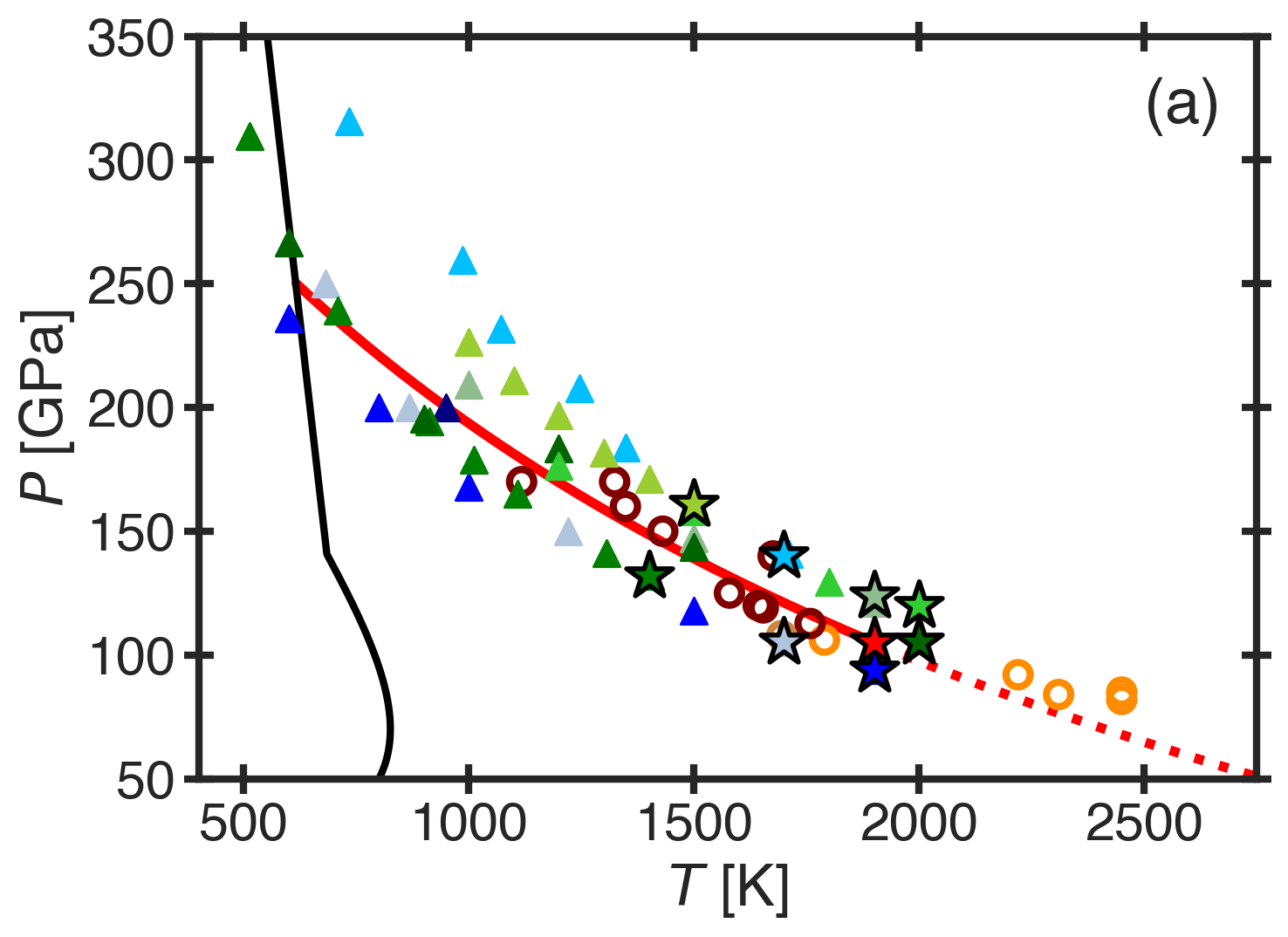}
    \includegraphics[width=\linewidth]{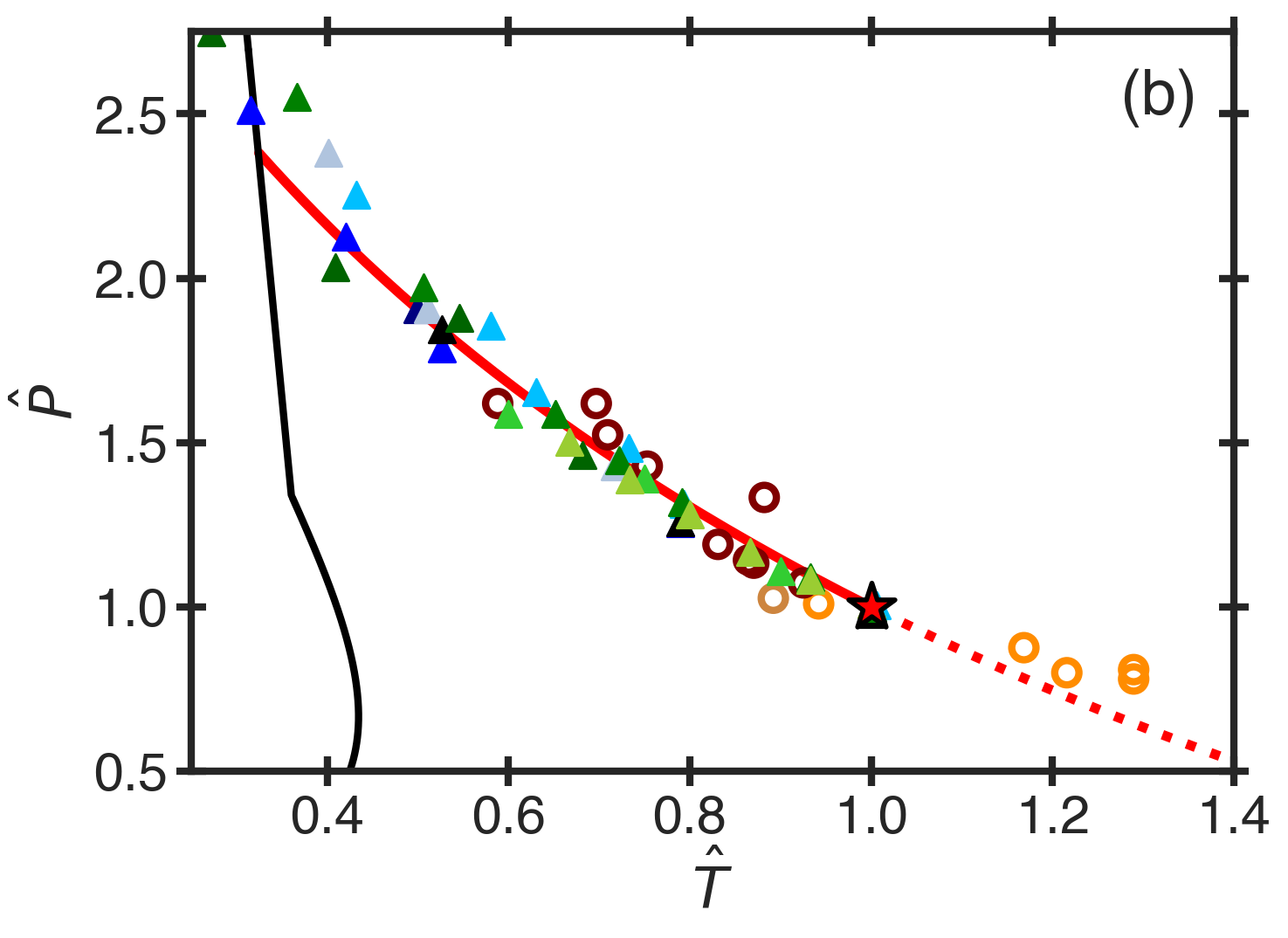}
    \caption{Unifying the different simulation results with experimental data of hydrogen by the generalized law of corresponding states. (a) Experimental and simulation results for the fluid-fluid phase transition (FFPT). (b) Unified representation of the FFPT by reducing pressure, $\hat{P}=P/P_\text{c}$, temperature, $\hat{T}=T/T_\text{c}$, and the critical value of the entropy, $\hat{S}=S/S_\text{c}$. In (a) and (b), the open circles are the experimental data of Zaghoo \textit{et al.} \cite{Zaghoo_H_2013,Zaghoo_H_2016,Zaghoo_H_2017} (dark brown), McWilliams \textit{et al.} \cite{McWilliams_H_2016} (light brown), and Ohta \textit{et al.} \cite{Ohta_H_2015} (orange). The computational results are indicated by the triangles: blue tints correspond to the Density Functional Theory (DFT) simulations of Bonev \textit{et al.} \cite{Bonev_MeltSim_2004} (dark blue), Morales \textit{et al.} \cite{Morales_H_2010} (blue), Hinz \textit{et al.} \cite{Heinz_H_2020} (sky blue), and Karasiev \textit{et al.} \cite{Cheng_Comment_2021} (light blue). Meanwhile, green tints correspond to the Quantum Monte Carlo (QMC) simulations of Morales \textit{et al.}\cite{Morales_H_2010} (dark sea green), Lorenzen \textit{et al.} \cite{Lorenzen_H_2010} (green), Perlioni \textit{et al.} \cite{Pierleoni_H_2016} (dark green), Mozzola \textit{et al.} \cite{Mazzola_H_2018} (lime green), and Tirelli \textit{et al.} \cite{Tirelli_MachineLearn_2022} (yellow green). The colored stars correspond to the reported (or adopted in this work) critical points for each data set. The solid black curve is the solid-fluid phase transition line as discussed in Fig.~\ref{Fig_PT}, and the red solid line is the FFPT predicted in this work.}
    \label{Fig_CorrespondStates}
\end{figure}

The significant discrepancy between the results of different computational models makes it impossible to utilize these results for a single equation of state. However, presenting the same results in reduced variables, as suggested by the law of corresponding states, allows the computational results to be used along with the experimental data for thermodynamic modeling. In Figure~\ref{Fig_CorrespondStates} all of the available computational and experimental data on the fluid-fluid phase transition are presented in real units of pressure and temperature (Fig.~\ref{Fig_CorrespondStates}a) and in reduced variables (Fig.~\ref{Fig_CorrespondStates}b), $\hat{P}=P/P_\text{c}$ and $\hat{T}=T/T_\text{c}$, where $P_\text{c}$ and $T_\text{c}$ are the critical pressures and temperatures obtained (or adopted) from different works. We found that the simulation data based on Quantum Monte Carlo (QMC) could also be collapsed into the universal phase diagram by reducing the entropy by the critical value of the entropy, $\hat{S} = S/S_\text{c}$. In classical thermodynamics, the reference value for the entropy is arbitrary. Commonly, the value, $S_\text{c}$, is adopted as $S_\text{c} = \text{d}\hat{P}/\text{d}\hat{T}|_{T=T_\text{c}}$ \cite{Gorodetskii_Asymmetry_1995,Wang_Asymmetry_2007,Uralcan_Water_2019}, which was found to be $S_\text{c} = 0.8$ for all QMC simulations.

Thermodynamically, the phenomenon of the fluid-fluid transition in hydrogen can be modeled through the interconversion reaction, $\ce{A <=> B}$, \cite{Anisimov_Polyamorphism_2018} where state A  represents the free atoms of hydrogen and state B represents dimerized hydrogen atoms. The total Gibbs energy per hydrogen atom (reduced by $RT_\text{c}$, where $R$ is the ideal-gas constant) is  
\begin{equation}
    G=G_{A}+x G_{BA}+G_\text{mix}(x)
\end{equation}
where $G_{BA} = G_B - G_A$, such that $G_A$ and $G_B$ are the Gibbs energies of hydrogen in the monatomic or diatomic states, respectively, $x$ is the fraction of hydrogen atoms in the diatomic state, and $G_\text{mix}$ is the Gibbs energy of mixing of these two alternative states. We model $G_\text{mix}$ as a sum of two parts: an asymmetric quasi-ideal mixing of diatomic and monatomic hydrogen and a non-ideal excess Gibbs energy of mixing in the form 
\begin{equation}\label{Eq_G_Mix}
    G_\text{mix}(x)=\hat{T}\left[ \frac{x}{2} \ln\frac{x}{2} +  (1-x) \ln{(1-x)}\right]+\omega(\hat{T},\hat{P}) \ x (1-x)
\end{equation}
We approximate the dimensionless non-ideality parameter, $\omega =\omega(\hat{T},\hat{P})$, up to first order in $\Delta \hat{T}$ and $\Delta\hat{P}$, as
\begin{equation}\label{Eq_Omega}
   \omega(T,P) = \omega_0 - \omega_T\Delta\hat{T} + \omega_P\Delta\hat{P}
\end{equation}
where $\Delta\hat{T} = \hat{T}-1$ and $\Delta\hat{P} = \hat{P}-1$.

\begin{figure}[t]
    \centering
    \includegraphics[width=\linewidth]{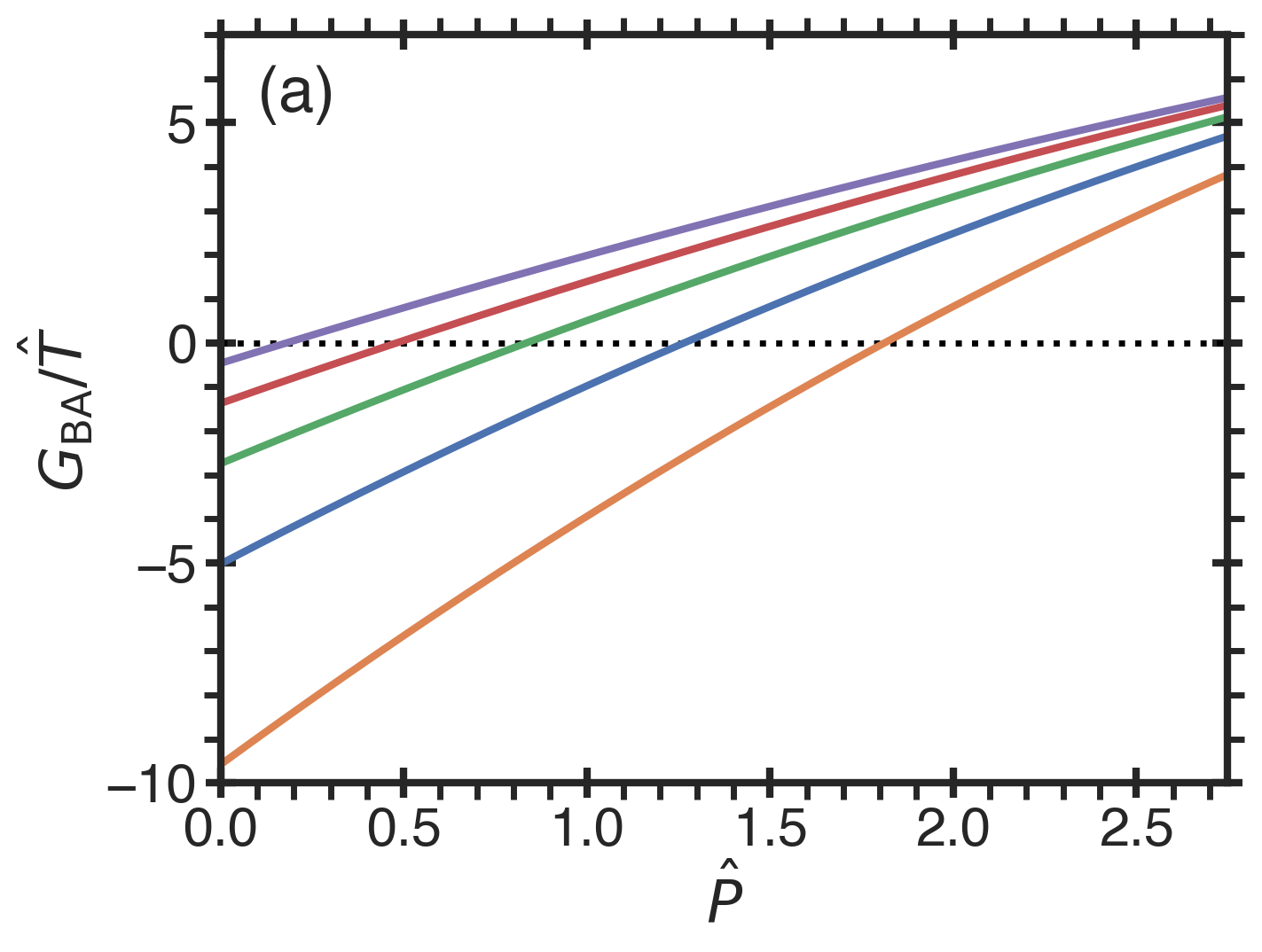}
    \includegraphics[width=\linewidth]{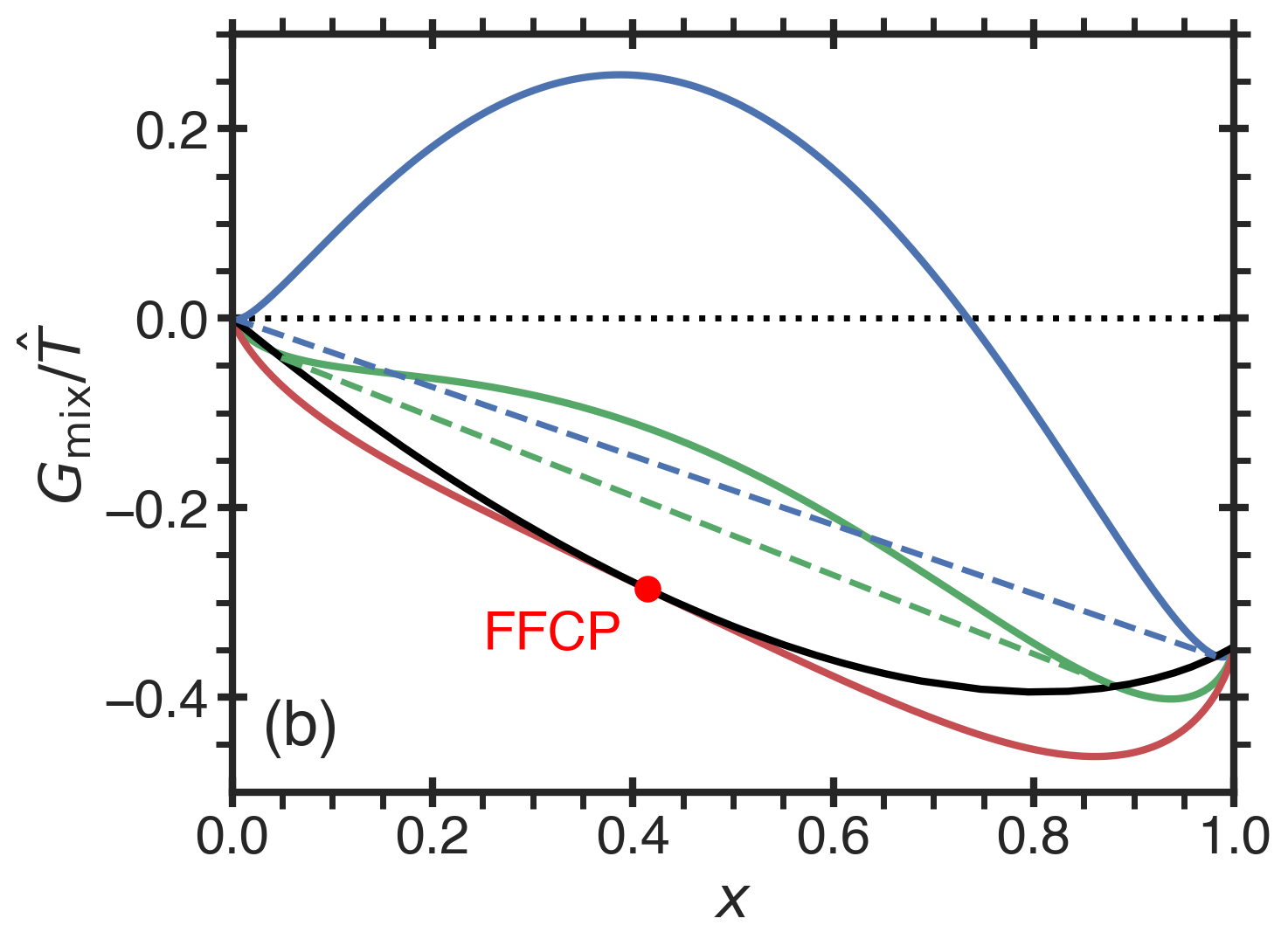}
    \caption{The components of the Gibbs energy (per atom) for hydrogen in the vicinity of the fluid-fluid phase transition. (a) The Gibbs energy of reaction, $G_\text{BA}$, as given by Eq.~(\ref{Eq_G_BA}). The isotherms are $T=0.5T_\text{c}$ (orange), $T=0.75T_\text{c}$ (blue), $T=T_\text{c}$ (green), $T=1.25T_\text{c}$ (red), and $T=1.5T_\text{c}$ (purple). (b) The Gibbs energy of mixing, $G_\text{mix}$, as given by  Eq.~(\ref{Eq_G_Mix}). $G_\text{mix}$ is shown as a function of the fraction of hydrogen atoms in the diatomic state, $x$, for isotherms $T=0.5T_\text{c}$ (blue), $T=0.75T_\text{c}$ (green), and $T=T_\text{c}$ (red) at $P=P_\text{c}$. The solid curve corresponds to the fluid-fluid coexistence.}
    \label{Fig_lnK}
\end{figure}

The FFCP parameters are determined from the thermodynamic stability criteria that $\partial^2 G/\partial x^2 = 0$ and $\partial^3 G/\partial x^3 = 0$, such that the critical fraction of hydrogen atoms is $x_\text{c} = \sqrt{2}-1$, and the critical temperature is $T_\text{c} = 2(2-\sqrt{2})^2\omega_0$. We note that the first study to apply the two-state thermodynamic approach to high-pressure hydrogen was presented by Cheng \textit{et al.}\cite{Cheng_H_2020}. While the predictions of Cheng \textit{et al.} for the FFPT are not in agreement with the results of all other simulations and experimental studies \cite{Cheng_Comment_2021}, their study provides a reasonable idea for how the non-ideality parameter, $\omega$, might depend on pressure and temperature. Based on the suggested trend, we optimized $\omega_T$ and $\omega_P$ to agree with the behavior of hydrogen from the available computational data \cite{Morales_H_2010,Pierleoni_H_2016,Mazzola_H_2018,Geng_H_2019,Heinz_H_2020,Tirelli_MachineLearn_2022}, and consequently, adopted these parameters as $\omega_T = 2.062$ and $\omega_P = -0.175$. The asymmetric Gibbs energy of mixing is illustrated in Fig.~\ref{Fig_lnK}a along with the fluid-fluid coexistence, calculated via the common tangent method, and the limit of absolute stability (spinodal), calculated via the thermodynamic stability conditions.

The condition for chemical-reaction equilibrium is given by $\partial G/\partial x=0$, resulting in the balance of the Gibbs energy of reaction, $G_\text{BA}$, and the exchange chemical potential of mixing, $\mu_\text{mix} =\partial G_\text{mix}/\partial x|_{\hat{T},\hat{P}}$, such that thermodynamic equilibrium follows from
\begin{equation}\label{Eq_rxn_cond}
   G_\text{BA}= -\mu_\text{mix}
\end{equation}
We approximate the Gibbs energy of reaction, $G_\text{BA} = G_\text{BA}(\hat{T},\hat{P})$, up to second order in $\hat{T}$  and $\hat{P}$, as
\begin{equation}\label{Eq_G_BA}
    G_\text{BA} = \epsilon - \alpha \hat{T} + \beta \hat{P} + \gamma \hat{T}\hat{P} + \frac{\delta}{2}\hat{T}^2 - \frac{\kappa}{2}\hat{P}^2 
\end{equation}
where $\epsilon$, $\alpha$, and $\beta$ are the energy, entropy, and volume changes of the reaction, while $\gamma$, $\delta$, and $\kappa$ are proportional to the volumetric expansivity, isobaric heat capacity, and isothermal compressibility changes of the reaction, respectively. To balance the Gibbs energy of reaction, Eq.~(\ref{Eq_G_BA}), with the derivative of the Gibbs energy of mixing, we express $G_\text{BA}$ as an expansion in $\Delta\hat{T}$ and $\Delta\hat{P}$ as 
\begin{equation}\label{Eq_cxc_cond}
   G_\text{BA} = u - a\Delta\hat{T}+ b\Delta\hat{P} + g\Delta\hat{T}\Delta\hat{P} + \frac{d}{2}(\Delta\hat{T})^2 - \frac{k}{2}(\Delta\hat{P})^2
\end{equation}
where the modified coefficients of the thermodynamic balance, Eq.~(\ref{Eq_cxc_cond}), are related to the coefficients of reaction, Eq.~(\ref{Eq_G_BA}), as:
\begin{equation}
\begin{split}
    \varepsilon &= u + a - b+ g + \frac{d}{2} - \frac{k}{2} \\
    \alpha &= a + g + d\\
    \beta &= b - g + k
\end{split}
\end{equation}
along with $\gamma = g$, $\delta = d$, and $\kappa = k$. 

If the Gibbs energy of mixing, $G_\text{mix}$, would be symmetric with respect to $x$, then $G_\text{BA}=-\mu_\text{mix} =0$, could describe the conditions for both reaction equilibrium and fluid-fluid phase equilibrium \cite{Anisimov_Polyamorphism_2018}. However, since the monatomic and diatomic mixing is asymmetric, the condition for the balance of phase and reaction equilibrium, Eq.~(\ref{Eq_rxn_cond}), is given through
\begin{equation}\label{Eq_Asym}
    \frac{\mu_\text{mix}}{\hat{T}} = a_2 \left(\frac{\omega(T,P)}{\hat{T}}-\omega_0\right)^2 + a_1\left(\frac{\omega(T,P)}{\hat{T}}-\omega_0\right)+a_0
\end{equation}
where the coefficients $a_0 = -0.502$, $a_1 = 0.166$, and $a_2 = -0.071$.

The developed equation of state is formulated through the Gibbs energy for the system as a function of temperature and pressure. Due to the interconverting nature, the two-states of hydrogen are thermodynamically equivalent to a single component system. Consequently, this produces an equation of state in terms of the equilibrium fraction of dimerized atoms, $x_e = x_e(T,P)$, and the density of the system, $\rho = \rho(T,P)$. Our equation of state contains seven adjustable parameters: five from the Gibbs energy of reaction, $G_\text{BA}$, {($u$, $a$, $b$, $g$, and $k$)}, Eq.~(\ref{Eq_G_BA}) and {two} from the non-ideality parameter in the Gibbs energy of mixing {($\omega_T$ and $\omega_P$)}, Eq.~(\ref{Eq_Omega}). We reduce the number of adjustable parameters from the following analysis of the available computational data on hydrogen in the vicinity of the fluid-fluid critical point.

From the computational heat capacity data presented by Karasiev \textit{et al.} \cite{Cheng_Comment_2021}, {we approximate the heat-capacity change of reaction} to be $\delta\approx 0$, and from the computational isothermal-compressibility data presented in the supplemental material of Pierleoni \textit{et al.} \cite{Pierleoni_H_2016}, we approximate $\kappa \approx \SI{0.625}{[\milli\meter^3/\giga\pascal\cdot\mole]}$. Additionally, we adopt $\epsilon = \SI{-108}{[\kilo\joule/\mole]}$ based on the known value of the bond dissociation energy of H$_2$ \cite{H_Diss_Energy}. As discussed above, we adopt $\omega_T = 2.062$ and $\omega_P = -0.175$. 

From these findings, we have reduced the number of free parameters to three: $a$, $b$, and $g$. We determined the values of the remaining free parameters as $a =  -4.95$, $b = 0.044$, and $g = 0.0124$ from the computational and experimental data utilizing the generalized law of corresponding states (Fig.~\ref{Fig_CorrespondStates}). Using the relations between these parameters and the physical parameters in Eq.~(\ref{Eq_G_BA}), we estimate: the entropy change of the reaction as $\alpha=\SI{-34.0}{[\joule/\kelvin\cdot\mole]}$, the volume change of the reaction $\beta=\SI{393}{[\milli\meter^3/\mole]}$, and the volume-expansivity change of the reaction $\gamma=\SI{0.0677}{[\milli\meter^3/\kelvin\cdot\mole]}$. The Gibbs energy change of reaction is shown in Fig.~\ref{Fig_lnK}b. It demonstrates that the pressure is the major factor in the behavior of $G_\text{BA}$.

\begin{figure}[b]
    \centering
    \includegraphics[width=\linewidth]{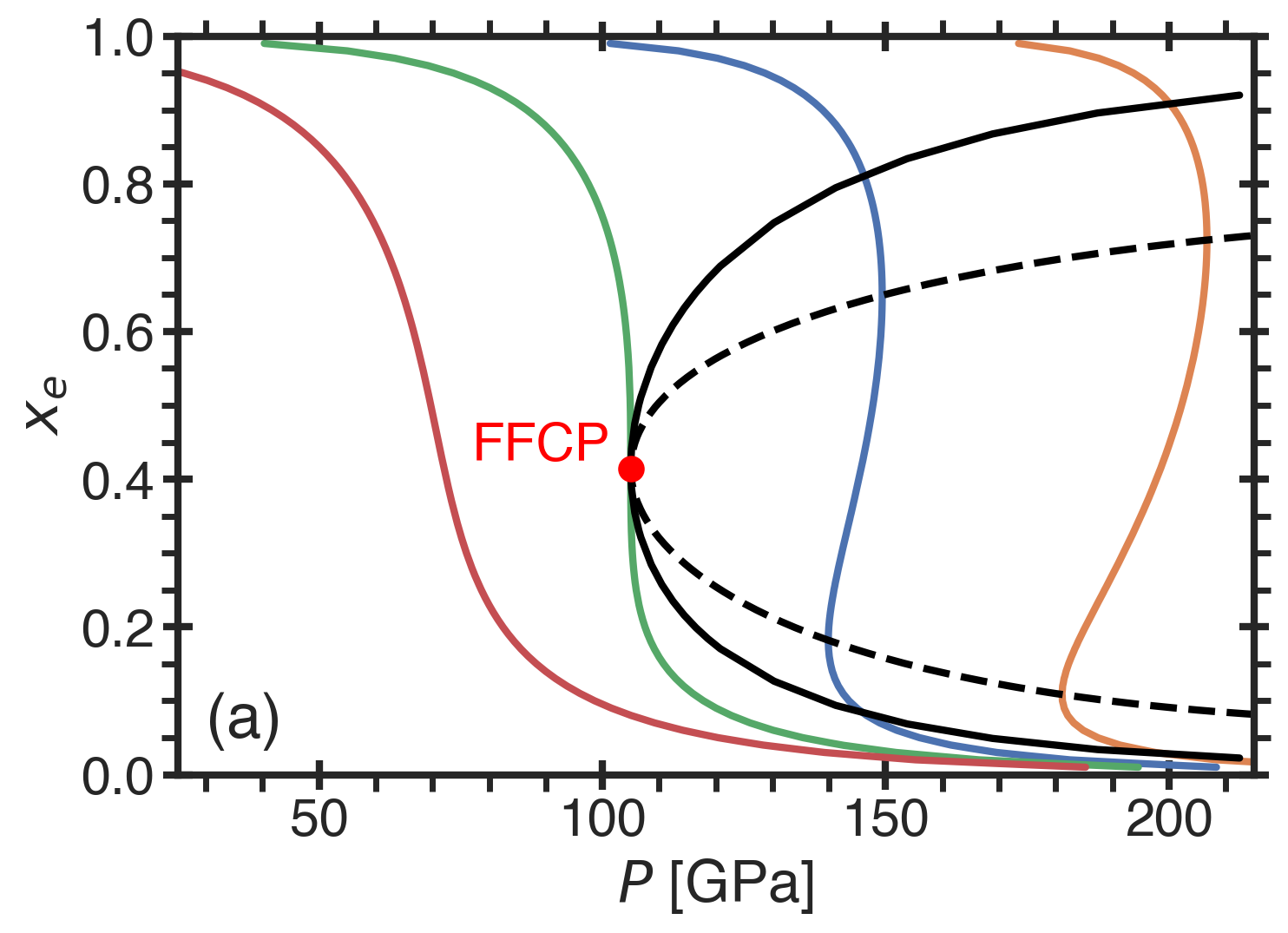}
    \includegraphics[width=\linewidth]{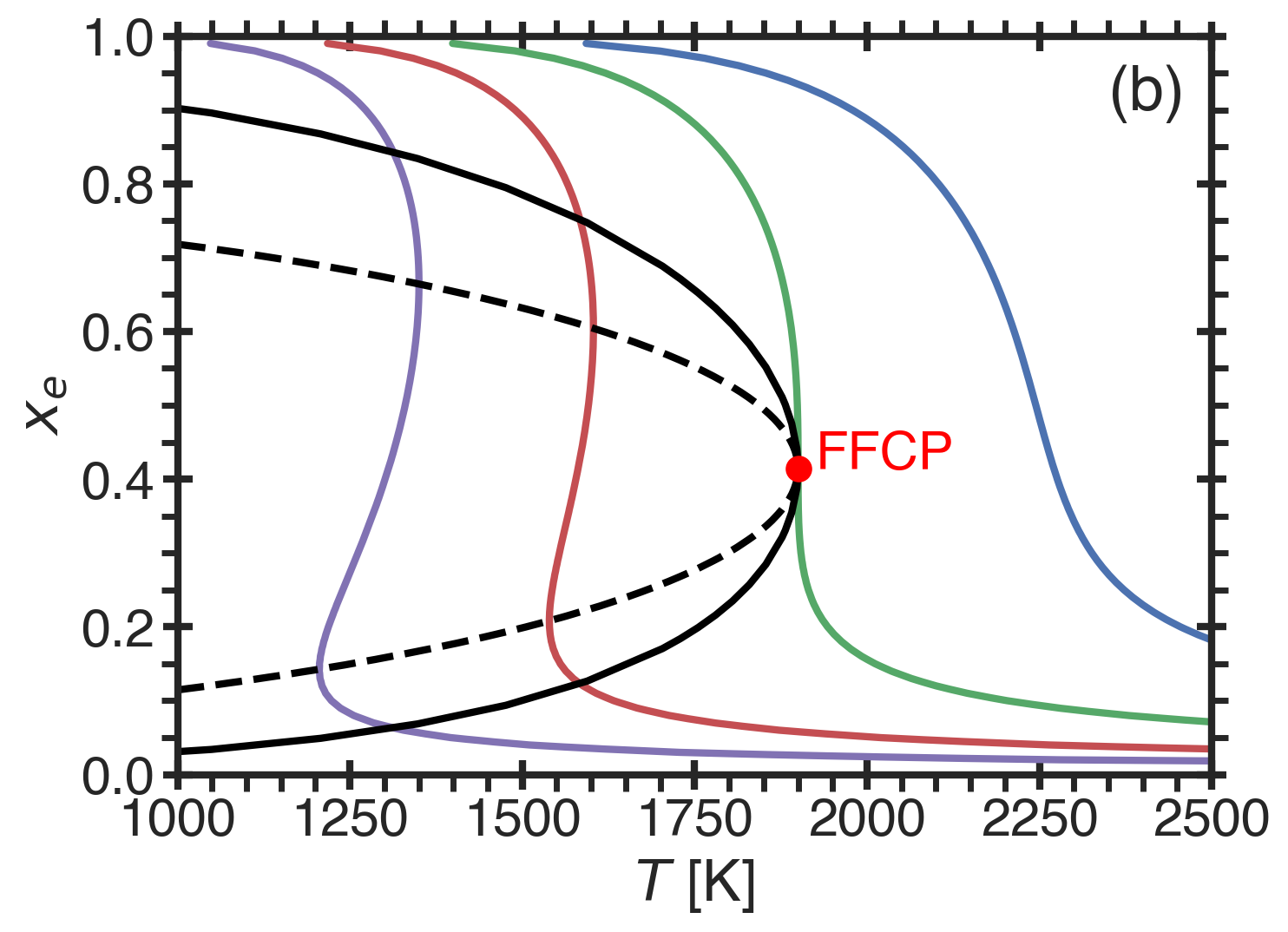}
    \caption{Equilibrium fraction of hydrogen atoms in the diatomic state, $x_e$. (a) Equilibrium fraction-pressure diagram for $T=0.5T_\text{c}$ (orange), $T=0.75T_\text{c}$ (blue), $T=T_\text{c}$ (green), and $T=1.25T_\text{c}$ (red). (b) Equilibrium fraction-temperature diagram for $P=0.75P_\text{c}$ (blue), $P=P_\text{c}$ (green), $P=1.25P_\text{c}$ (red), $P=1.5P_\text{c}$ (purple). The solid and dashed black curves are, respectively, the fluid-fluid coexistence and the limit of thermodynamic stability (spinodal).}
    \label{Fig_PTvsRho}
\end{figure}

Using the Gibbs energy of mixing, Eq.~(\ref{Eq_G_Mix}), the Gibbs energy of reaction, Eq~(\ref{Eq_G_BA}), and the variables determined from the {universal phase diagram}, the equilibrium fraction of hydrogen atoms in the dimerized state, $x_e$, is determined from Eq.~(\ref{Eq_rxn_cond}). The corresponding equilibrium-fraction phase diagrams are presented in Fig.~\ref{Fig_PTvsRho}(a,b). At higher temperatures and lower pressures, the equilibrium composition changes from the dimeric state $x_e = 1$ to the monomeric state $x_e = 0$.

\begin{figure}[t]
    \centering
    \includegraphics[width=\linewidth]{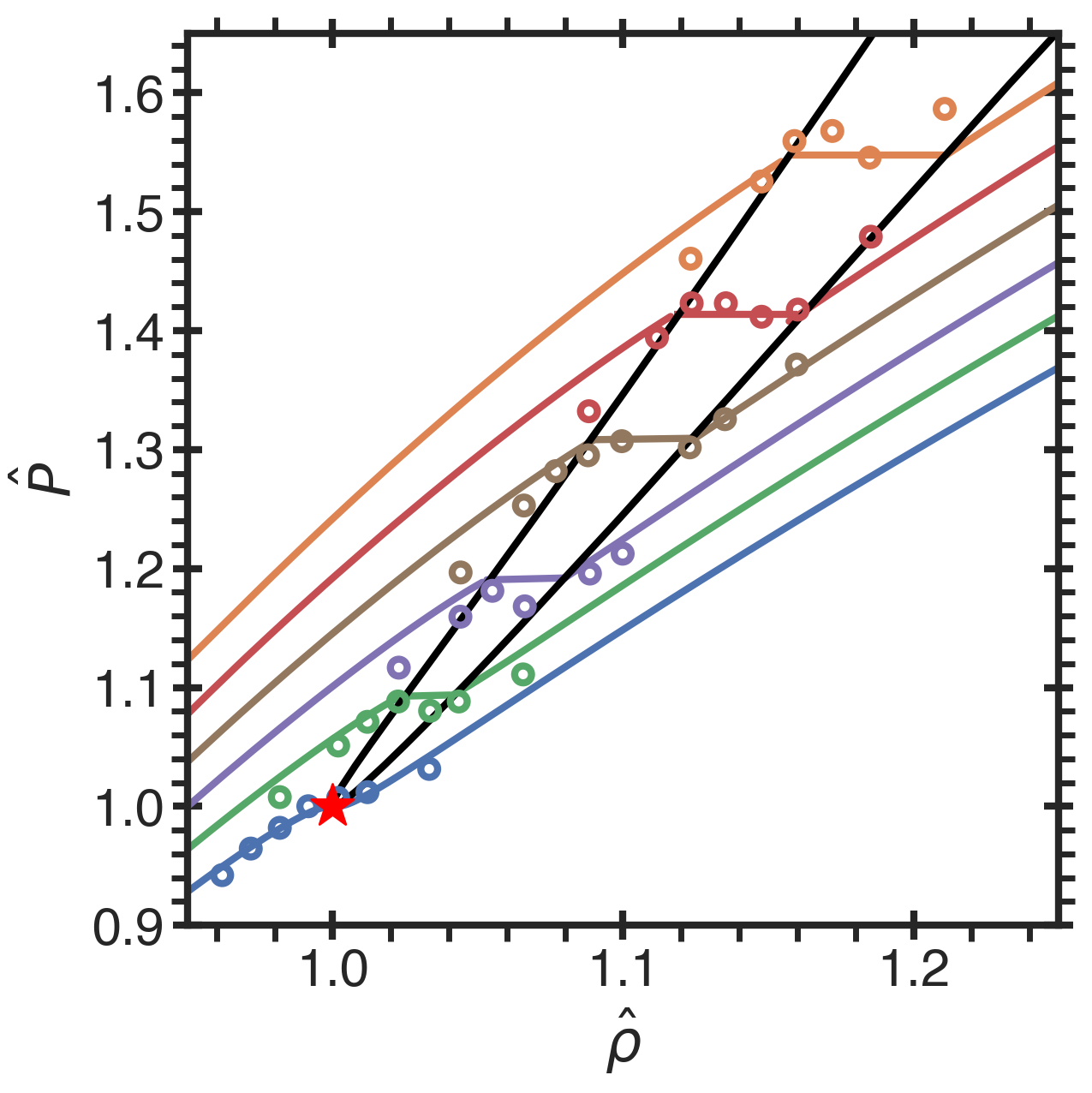}
    \caption{The pressure-density phase diagram of hydrogen based on the equation of state developed in this work. The open circles correspond to the QMC simulations of Tirelli \textit{et al.}\cite{Tirelli_MachineLearn_2022}. Isotherms are $\hat{T}=0.67$ (orange), $\hat{T}=0.73$ (red), $\hat{T} = 0.8$ (brown), $\hat{T}=0.87$ (purple), $\hat{T}=0.93$ (green), and $\hat{T}=1.0$ (blue). The fluid-fluid coexistence is shown by the solid black curve. The red star is the FFCP adopted in this work.}
    \label{Fig_Density}
\end{figure}

The density of species is expressed through the equilibrium fraction via \cite{Anisimov_Polyamorphism_2018}
\begin{equation}
    \hat{\rho}(\hat{P},\hat{T}) = \left(\frac{\partial G}{\partial \hat{P}}\right)^{-1}_{\hat{T}} = \left[\frac{1}{\hat{\rho}_A} + x_e\frac{\partial G_{BA}}{\partial \hat{P}} + \frac{\partial\omega}{\partial \hat{P}} x_e \left(1-x_e\right)\right]^{-1}
\end{equation}
where $\rho_\text{A} = \rho_\text{A}(\hat{P},\hat{T})$ is the volume of the monatomic hydrogen state, and may be expressed to second-order in $\Delta\hat{T}$ and $\Delta\hat{P}$ as
\begin{equation}\label{Eq_RhoA}
    \hat{\rho}_A = \hat{\rho}_\text{c} - \hat{\rho}_0\Delta\hat{T} + \hat{\rho}_1\Delta\hat{P} + \hat{\rho}_2\Delta\hat{T}\Delta\hat{P} - \hat{\rho}_3\left(\Delta\hat{P}\right)^2 + \hat{\rho}_4\left(\Delta\hat{T}\right)^2
\end{equation}
Using the most recent QMC simulations presented in Tirelli \textit{et al.} \cite{Tirelli_MachineLearn_2022}, $\rho_A$ is estimated by Eq.~(\ref{Eq_RhoA}) with coefficients: $\hat{\rho}_\text{c}=1.01$, $\hat{\rho}_0 = 0.25$, $\hat{\rho}_1 = 0.56$, $\hat{\rho}_2=0.56$, $\hat{\rho}_3 = 0.21$, and $\hat{\rho}_4 = 0.12$. The corresponding pressure-density phase diagram is presented in Fig.~\ref{Fig_Density}, and demonstrates a good agreement with the computational data in the vicinity of the FFCP.

We note that the properties observed in experimental studies (\textit{e.g.} conductivity, reflectivity, thermal efficiency, etc.) could be indirectly related to the proper order parameter for the FFPT in hydrogen. In the thermodynamic scheme presented in this work, the corresponding order parameter is the difference between the fraction of dimerization and its critical value, $x - x_\text{c}$. The measureable quantities (such as density or conductivity) are  coupled to the order parameter.

The rate of dimerization/dissociation could also affect the observation of the FFPT in hydrogen. Recent simulations by Geng \textit{et al.} indicate that the interconversion between H$_2$ and H is fast compared to the self-diffusion of species \cite{Geng_H_2019}. This effect of interconversion could produce the phenomenon of phase amplification, the growth of one phase at the expense of the other \cite{Shum_Phase_2021,Longo_PT_2022}. Phase amplification occurs to avoid the formation of an energetically unfavorable interface between alternative stable phase domains. In macroscopic systems, where the interfacial energy is much smaller than the bulk energy, the formation of a metastable interface becomes less unfavorable, and the possibility that the system would form an interface drastically increases \cite{Longo_PT_2022}. Lastly, depending on the simulation conditions, due to a non-zero volume of the dimerization reaction, phase amplification may or may not occur depending on the simulation ensemble \cite{Longo_PT_2022}. These factors could contribute to the challenge in observing the FFPT in hydrogen.

In conclusion, hydrogen at extreme conditions is an example of a polyamorphic fluid. There is a remarkable analogy between the challenges in thermodynamically modeling the fluid-fluid phase transition in hydrogen and other polyamorphic substances, such as supercooled water. In this work, we have outlined the steps to thermodynamically model the FFPT in hydrogen. Using the most recent computational and experimental studies \cite{Zaghoo_H_2013,Zaghoo_H_2016,Zaghoo_H_2017,McWilliams_H_2016,Ohta_H_2015,Morales_H_2010,Lorenzen_H_2010,Morales_Review_2012,Xu_Supercritical_2015,Pierleoni_H_2016,Mazzola_H_2018,Geng_H_2019,Heinz_H_2020,Cheng_H_2020,Cheng_Comment_2021}, we provide the first attempt to develop the equation of state for high-pressure hydrogen near the FFPT. We demonstrate that by using a generalized law of corresponding states (via reducing the pressure, temperature, and entropy by their critical values), the results of simulations can be reconciled. We also provide estimates of the entropy, volume, and volume-expansivity change of the reaction. 

In its current form, our equation of state has been optimized in the vicinity of the fluid-fluid critical point, but in the future, the proposed thermodynamic scheme could be refined upon the arrival of more comprehensive experimental and computational data  for hydrogen at extreme conditions. In particular, with more accurate estimates of the heat and volume change of the transitions from the solid-hydrogen phase to the alternative coexisting fluid phases, it could be possible to predict the change in the slope of the melting curve at the SFF-triple point. In addition, it would be desirable to investigate the dynamics of phase growth and its relation with the rate of dimerization in high-pressure hydrogen.

\acknowledgements{We thank Sergey Buldryev, Avik Dutt, and Nikolay Shumovskyi for useful discussions. We also acknowledge Genri Norman and Ilnur Saitov for bringing several key references to our attention.} This work is a part of the research collaboration between the University of Maryland, Princeton University, Boston University, and Arizona State University supported by the National Science Foundation. The research of N.R.F. was supported by the UMD Chemical Physics Program. The research of T.J.L. and M.A.A. was supported by NSF award no. 1856479.

\section*{Author Declarations}
\subsection*{Conflict of Interest}
The authors have no conflicts to disclose.

\section*{Data Availability}
Data sharing is not applicable to this article as no unpublished data were created or analyzed in this study.

\section*{References}
\bibliography{ref}

\end{document}